\documentclass[11pt]{amsart}
\usepackage{geometry}                
\usepackage{graphicx}
\usepackage{amssymb}
\usepackage{epstopdf}
\usepackage{amssymb}
\usepackage{yfonts}

\DeclareMathAlphabet{\mathpzc}{OT1}{pzc}{m}{it}
\usepackage{bm}
\usepackage{mathrsfs}
\usepackage{hyperref}
\usepackage[toc,page]{appendix}

\DeclareMathAlphabet{\mathpzc}{OT1}{pzc}{m}{it}

\renewcommand {\P}{\mathcal{P}}

\newcommand {\K}{\mathcal{K}}

\newcommand {\A}{\mathcal{A}}
\renewcommand {\H}{\mathscr{H}}
\renewcommand {\L}{\mathcal{L}}

\newcommand {\U}{\mathrm{U}}

\newcommand {\G}{\mathcal{G}}

\newcommand {\M}{\mathcal{M}}

\newcommand{\g}{\mathfrak{g}}

\newcommand{\tg}{\textgoth{g}}
\newcommand{\tk}{\textgoth{k}}

\title{A new approach to superstring}

\author {  A. Schwarz\\ Department of Mathematics\\ 
University of 
California \\ Davis, CA 95616, USA,\\ schwarz @math.ucdavis.edu}

\begin{document}
\maketitle
\begin{abstract} {We show how starting with one-string space of states in BRST formalism one can construct a large class of physical quantities containing, in particular, scattering amplitudes for bosonic string and superstring. The same techniques work for heterotic string.}
\end{abstract}
\section {Introduction}
In the paper \cite{SCH} I formulated a new approach to string theory. It was based on some general considerations that could be used for any quantum theory obtained from classical theory with constraints generating a Lie algebra with respect to the Poisson bracket.
This approach was applied to the bosonic string. It led to the conclusion that the one-string space of states contains all information that is necessary to calculate the scattering of interacting strings. This is not very surprising because in some sense the one-string space contains all $n$-string states (this follows from the remark that for any two-dimensional conformal field theory, a conformal manifold with $n$ incoming boundaries and one outgoing boundary specifies a map $\mathcal{ H}^n\to \mathcal {H}$  where $\mathcal {H}$ stands for the space of states). The fact that the one-string space of states contains the information about interactions is important because it opens the way for a formulation of string theory in terms of algebraic and geometric approaches  \cite {GA1},\cite {GA}, \cite{GA2}, \cite{GA3}.

It seems that by using the algebraic and/or geometric approaches one can obtain the expressions for string amplitudes from the first principles, but at this moment we should use comparison with well-known results to 
 check that the expressions given in \cite {SCH} and in the present paper give string amplitudes. (We compare our calculations with \cite{BE}, \cite{BEL}, \cite{BELO}; it is more difficult to compare them with  \cite{AGG} or with \cite {WIT}).

The present paper is a follow-up paper to \cite {SCH}. I slightly modified the approach of \cite {SCH} to analyze superstring; the same modification should work for heterotic string. The exposition is completely independent of \cite {SCH} and does not depend on string theory textbooks (at least formally). Section 3 contains a review of Segal's definition of CFT \cite{SEG} and of operator formalism in string theory \cite{A},\cite{AGG}, however, this review is very short, therefore the familiarity with Segal's definition and operator formalism would be very helpful  ( as well as familiarity with \cite{SCH}). We give also definitions of
superconformal field theory (SCFT), of topological conformal field theory (TCFT)
and topological superconformal field theory (TSCFT). Our definitions of TCFT and TSCFT are not standard, this modification allows us to simplify proofs.

Let me give a short description of my approach.

For every super Lie algebra $\tg$ the direct sum of vector spaces $\tg$ and $\Pi \tg$ has a structure of differential super Lie algebra denoted by $\tg'.$ ( Here $\Pi$ stands for parity reversion.) Sometimes it is convenient to extend $\tg'$ considering the differential  as an element of extended super Lie algebra $\tg''.$
We will apply these constructions to the Lie algebra diff of complex vector fields on a circle and to supersymmetric generalization of diff: to the direct sum $W$ of super Lie algebra $W_{NS}$ of vector fields on Neveu-Schwarz (NS) supercircle and super Lie algebra $W_R$ of vector fields on Ramond (R) supercircle.  Let us denote by $\mathcal{E}'$ the one-string space of states of string theory in BRST formalism. Then for some choice of super Lie algebra $\tg$ the super Lie algebra $\tg''$ acts on $\mathcal{E}'$ (and, more generally, the direct sum of $n$ copies of $\tg''$ acts on $n$-th tensor power of $\mathcal{E}'$). In the case of open bosonic string the Lie algebra diff$''$ acts on one-string space, in the case of closed bosonic string we have an action of a direct sum of two copies of diff$''$, for the heterotic string we have an action of the direct sum of diff$''$ and $W''$ and for type II superstring we have an action of two copies of $W.''$ The action of the Lie algebra  $\textgoth{g}'$ on $\mathcal {E}'$ specifies a homomorphism $\psi$ of  $\textgoth{g}'$ into the space $\L$ of linear operators acting in $\mathcal{E}'$. (The space $\L$ is considered as super Lie algebra, and the homomorphism $\psi$ commutes with differentials. The differential in $\L$ is induced by BRST operator.) We assume that there exists a group (or semigroup) $\G$ such that the Lie algebra of $\G$ coincides with  $\textgoth{g}$ (or with a subalgebra of this algebra). Further, we assume that there exists a homomorphism $\Psi:\G'\to \L$ where $\L$ is regarded as a semigroup such that  $\Psi$ induces  $\psi$ as a homomorphism of the Lie algebra of $\G'$. In other words, {\it we assume that the homomorphism 
$\psi$ of Lie algebras can be integrated (exponentiated) into a homomorphism $\Psi$ of (semi)groups}. (We use the notation $\G'$ for $\Pi T\G$ considered as a supergroup.)
If $\sigma$ is a BRST-closed element of $\L$ then $\Psi^*(\sigma)$ is a function on $\G'$ that can be regarded as a closed pseudodifferential form on $\G.$ One can try to obtain physical quantities integrating this form over cycles in $\G$. Unfortunately, this idea does not lead to interesting quantities. However, it is possible to find a large class of Lie subalgebras $\tk \subset \tg$ such that $\Psi^*(\sigma)$ descends to a closed pseudo differential form on  the quotient
$\G/\tk$; sometimes this form descends even further. Integrating the form we obtained over some cycles
 in quotient space we can get interesting physical quantities; in particular, we can get string amplitudes for all types of strings in the framework of perturbation theory. 

More precisely, to get string amplitudes we should consider an action of the direct sum of $n$ copies of $\tg''$ on $(\mathcal{E}')^n.$ To define the subalgebra $\tk  \subset\tg+...+\tg$ in the case when $\tg=$diff  we take a one-dimensional complex manifold with $n$ embedded disks and consider complex vector fields on the boundaries of the disks that can be holomorphically extended to the complement of the disks. The definition of $\tk$ in other cases is similar.  

Our constructions can be modified in many ways. Our starting point was the one-string space of states. It seems that it is useful to represent states by L-functionals \cite{L}, \cite {FSCH}. The one-string state depends on the choice of the picture in $(\beta,\gamma)$-model;  L-functionals describe states in all pictures (and more). One can conjecture that the language 
of L-functionals can be used to replace to some extent the language of string field theory.

The construction of the subalgebra $\tk$ that we are using to get string amplitudes is related to (super)conformal theory. However, it is possible to apply our consideration to other subalgebras; it seems that using $\tk$ coming from infinite-dimensional Grssmannian one can obtain non-perturbative physical quantities.

\section{Preliminaries}

Let us review some basic definitions and constructions (see\cite {SC} or \cite {KSCH} for the definition of superspace we are using). 

We will extensively use the following construction. For every superspace $M$ we define a superspace $M'$  as a space of maps of $(0|1)$-dimensional superspace  $\mathbb{R}^{0|1}$ into $M.$ (Usually $M'$ is denoted by $\Pi TM$ because it can be interpreted as the total space $TM$ of the tangent bundle with inversed parity of the fibers; $\Pi$ stands for parity reversal.) The space $M'$ is equipped with a homological vector field $q$
(with an odd vector field  that specifies a first order differential operator $q$  obeying $q^2=0$.) This follows immediately from the fact that    $\mathbb{R}^{0|1}$  is equipped with a homological vector field.    This construction is functorial; this means that a map $M_1\to M_2$ induces a map $M'_1\to M'_2$ that agrees with homological vector fields. It follows from functoriality that for super Lie algebra $M$  the superspace $M'$  has a structure of differential super Lie algebra. If $M$ is a supergroup, $M'$ can be considered as a supergroup. A similar statement is true for super semigroups.

By definition, a function on $M'$ is a pseudodifferential form on $M$, the homological vector field $q$ on $M'$ corresponds to the de Rham differential. We can talk about closed and exact pseudodifferential forms (in the language of functions on $M'$  we say $f$ is closed if $qf=0$; it is exact if there exists $g$ such that $f=qg)$.

We will work mostly with infinite-dimensional objects. In this situation, there exist different versions of 
familiar definitions. Some finite-dimensional statements are wrong in infinite-dimensional case. It is essential for us that one can define a notion of Lie algebra of a semigroup (moreover, there are several versions of this definition,  but we disregard these subtleties). An action of a semigroup generates an action of corresponding Lie algebra in the same space.
Some of Lie algebras we consider ( for example, the algebra of complex vector fields on a circle) do not correspond to any group, but they can be regarded as Lie algebras of semigroups.

The statements above are correct also for super Lie algebras, supergroups, etc.
In what follows we sometimes omit the prefix "super" as in the previous paragraph.

 We will use the following important statement. {\it If $\tk$ is a Lie algebra acting on $X$ then $\tk'$ acts on $X'$ and $X'/\tk'$ can be identified with $(X/\tk)'$.}  
 
 The space $X/\tk$ can be defined as the space of leaves of the foliation generated by the action of $\tk$ on $X$. ( The action of $\tk$ specifies a map $\tau_m$ of $\tk$ to the tangent space of $X$ for every point $m\in X.$ Images of the maps $\tau_m$ specify a foliation. If $\tk$ is a Lie algebra of a connected group $K$ acting on $X$ then the foliation consists of tangent spaces to orbits of $K$  acting on $X$ and the leaves are orbits. This means that   $X/\tk$  can be identified with $X/K$.) 
 
 If $\G$ is a a group or a semigroup and $\tk$ is a subalgebra of its Lie algebra $\tg$ we define 
 $\G/\tk$ 
as the space of leaves of the foliation generated by the action of $\tk$ on $\G$ from the right. 
 Alternatively one
 can define the quotient $\G/\tk$  of semigroup $\G$ with respect to a subalgebra $\tk$ of its Lie algebra $\tg$ as a connected space  $M$
 where $\G$ acts transitively from the left with Lie stabilizer $\tk$ at one of points of $M$.  (We say that $\G$ acts transitively on $M$ if it induces a surjective map $\tau_m$ of the Lie algebra $\g$ to the tangent space of  $M$ at any point $m\in M.$ The Lie stabilizer at the point $m$ is defined as a kernel of $\tau_m.$) 
 Notice that the alternative definition is less rigorous. It does not define the quotient unambiguously.  However, it is very convenient; we use it.

If $M$ is an $(m|n)$-dimensional supermanifold with local coordinates $(x^1,...,x^m, \theta^1,...,\theta^n)$ then $M'$ can be identified with $(m+n|m+n)$-dimemensional supermanifold $\Pi TM$ with local coordinates $(x^1,...,x^m, \theta^1,...,\theta^n, dx^1,...,dx^m, d\theta^1,...,d\theta^n).$ Here $x^i,d\theta^j$ are even coordinates, $\theta^j, dx^i$ are odd coordinates. 

 Functions on $M'=\Pi TM$ can be regarded as pseudodifferential forms on $M$ ( if $M$ is an ordinary manifold they are inhomogeneous differential forms on $M$). We can integrate them over $M$ by doing the conventional integral with respect to even variables and the Berezin integral with respect to odd variables; the answer does not depend on the choice of coordinates in $M$. (Notice that for ordinary manifold $M$ the answer depends only on the highest degree term in inhomogeneous differential form.) We will 
need an integral of the pseudodifferential form
\begin {equation} \label{EXX} F(x,\theta)e^{K_{\alpha}(x,\theta)d\theta^{\alpha}}e^{L_i(x,\theta)dx^i}\end{equation}
where $K_{\alpha}$ are even functions and $L_i$ are odd functions.
Doing the integral with respect to $d\theta^{\alpha}$ and to $dx^i$ we see that the integral we need is equal to the integral with respect to $x,\theta$ of
\begin{equation}\label{INT}
F(x,\theta)\delta(K_1(x,\theta))...\delta(K_n(x,\theta))L_1(x,\theta)...L_m(x,\theta).
\end{equation}
Notice, that for odd $L$ we have $\delta(L)=L$, hence we in (\ref{INT}) we can write $\delta(L_i)$ istead of $L_i$.

If  $N\subset M$ then $N'\subset M'$. This means that a pseudofifferential form on $M$ (=a function on $M'$) can be regarded as a pseudodifferential form on $N$ (= a function on $N' $).  It follows that every pseododifferential form on (finite-dimensional or infinite-dimensional)  (super)space $M$ can be integrated over any finite-dimensional (super)space $N\subset M$ (one should restrict it to $N$ and integrate).
However, the integral over $N$ can diverge. Notice, that in the case when the pseudodifferential form on $M$ can be represented by the formula (\ref {EXX}) the same is true for its restriction to $N$, hence to calculate the integral over $N$ we can apply (\ref {INT}).

One can prove an analog of Stokes' theorem for the integration of pseudodifferential forms.

If $\tg$ is a super Lie algebra then $\tg'$  is a differential super Lie algebra. As a vector space it can be represented as a direct sum 
$\tg\oplus \Pi\tg$; the operation in the first summand coincides with the operation in $\tg$, the 
commutator of an element of the first summand and an element of the second summand lies in the second summand (it is specified by the action of $\tg$ on $\Pi\tg$) and, finally, two elements of the second summand super commute.  The differential $q$ maps $\tg$ onto $\Pi\tg.$ One can consider a larger Lie algebra  $\tg''$ adding to $\tg'$ the differential $q$ as an element.

 If $\textgoth {g}$ is a  Lie algebra  with generators $T_k$ and  commutation relations
 $[T_k,T_l]=f_{kl}^rT_r$ the Lie superalgebra $\textgoth {g}'$ has even generators $T_k$, odd generators $b_k$ and commutation relations  $[T_k,T_l]=f_{kl}^rT_r$, $[T_k,b_l]=f_{kl}^rb_r$, $[b_k,b_l]_+=0.$  To get $\tg''$ we should add an odd  generator  $q$ (the differential) obeying $[q,q]_+=0, [q,b_i]_+=T_i,[ q,T_i]=0.$.
 
  Sometimes it is convenient to use the Lie algebra
 that can be obtained as a semidirect product of $\tg'$ and the Lie algebra of vector fields on
 $\mathbb{R}^{0|1}$. The latter Lie algebra is generated by two elements (an odd element $q$ and an even element $n$); it acts on $\tg'$ because $\tg'$ can be regarded as a space of maps from  $\mathbb{R}^{0|1}$ to $\tg.$ The generator $n$ can be interpreted as a ghost number. 
 
 In what follows we could consider this larger algebra instead of $\tg''.$

  If a group $\G$ has a Lie algebra $\tg$ then any element  $g'$of  the group $\G'$ can be represented in the form
\begin{equation}\label{GGG}
g'=\exp(b)g
\end{equation}
where $g\in \G, b\in\Pi\tg$ and $\exp$ stands for the exponential map $\tg'\to \G'.$

We denote by $\mathcal {E}'$ the space of states of quantum theory in BRST-formalism. 

{\it Let us assume that a super Lie algebra $\textgoth{g}'$ acts on  $\mathcal {E}'$ and the  differential $q$ on 
$\textgoth {g}'$ is compatible with the BRST differential $Q.$}

In other words, we assume that there exists a Lie algebra homomorphism $\psi$ of $\textgoth{g}'$ into the Lie algebra $\L$ of linear operators in   $\mathcal {E}'$ and $\psi q=Q\psi$ (the BRST operator acts in the space  $\L$ of linear operators as (anti)commutator).

Usually  $\mathcal {E}'$ can be regarded as a tensor product of the space $\mathcal{E}$ equipped with an action of a Lie algebra $\textgoth{g}$ by the space of ghosts $\mathcal{E}_{gh}$. (This picture corresponds to the quantization of classical theory with constraints generating a Lie algebra $\tg$ with respect to the Poisson bracket. If $T_i$ are generators of $\tg$ ghosts $b_i,c_i$ have opposite parity and satisfy canonical (anti)commutation relations. The space of ghosts is defined as a representation of $b_i, c_i$. (In the case of infinite number of degrees of freedom there exist inequivalent representations of CAR/CCR, hence this definition   is ambiguous; we should specify the representation we are using.)  The representations of $\tg$ in  $\mathcal{E}$ and in $\mathcal{E}_{gh}$ can be projective but we assume that they generate a genuine representation of $\tg$ in  $\mathcal {E}'$. (This corresponds to critical string in string theory.)  Iit is important to notice that our assumptions are also satisfied when $\mathcal {E}'$ is a one-string space of states in BRST formalism for any version of critical (super)string theory. ( In this case
$\textgoth {g}$ is the Lie algebra  diff of complex vector fields on the circle $S^1$ for open bosonic string, a direct sum of
two copies of diff for closed bosonic string, and some extension of diff$\oplus$diff for superstring and heterotic string.)
More generally, our assumptions are satisfied  for topological quantum field theories of Witten type (recall that the target of (super)string theory
can be regarded as topological (super)conformal field theory).

Let us assume that $\tg$ is a Lie algebra of semigroup $\G$. Then $\tg'$ is a Lie algebra of semigroup $\G '$ and the differential in the semigroup $\G'$ agrees with the differential in $\tg'.$
{\it  We suppose that the homomorphism $\psi:\tg'\to\L$ where $\L$ is regarded as a Lie algebra comes from a homomorphism $\Psi:\G'\to \L$ where $\L$ is considered as a semigroup of linear operators acting in $\mathcal{E}'.$} (We are using the fact that a homomorphism of semigroups induces a homomorphism of corresponding Lie algebras.) The homomorphism $\Psi$ induces a map $\Psi^*$ of $\L^*$ into the space of functions on $\G'$ (into the space of pseudodifferential forms on $\G$). (Here $\L^*$ stands for the space of linear functionals on $\L.$) The map $\Psi^*$ agrees with differentials,
 hence for every BRST-closed element $\sigma$ of $\L^*$ we obtain a closed pseudodifferential form on $\G$.
 (We use the fact that the BRST operator $Q$ on $\L$ induces the BRST operator on $\L^*$ denoted by the same symbol.)
 
 One can try to construct physical quantities integrating $\Psi^*(\sigma)$ over cycles in $\G$. However, to obtain interesting physical quantities we should generalize this construction: we should fix a Lie subalgebra $\tk$ of $\tg$, assume that the closed form $\Psi^*(\sigma)$ descends to a closed pseudodifferential form on $\G/\tk$ and integrate this form over a cycle in the space $\G/\tk$.  
 
 The following construction gives a large class of examples where our conditions are satisfied.
 Specify a $Q$-closed element $\sigma\in \L^*$ by the formula
 \begin{equation}\label{SIG}
 \sigma (A)=\langle \chi|A|\rho\rangle.
 \end{equation}
 where $|\rho\rangle\in \mathcal{E}'$ and $\langle\chi|\in {\mathcal{E}'}^*$ are $Q$-closed.
 Then  a closed pseudodifferential form $\Psi^*(\sigma)$ on $\G$ considered as a function on $\G'$ is given by the 
 formula 
 \begin{equation}\label {PS}
 (\Psi^*\sigma)(g')=\langle\chi|\Psi(g')|\rho\rangle.
 \end{equation}
{\it  Impose the condition} $|\psi(\tk')\rho\rangle=0$ (In other words, we assume that $|\psi(k')\rho\rangle=0$ 
where $k'\in \tk'$  and $\tk$ stands for a Lie subalgebra of
 $\tg$. We say that $|\rho\rangle$ is $\tk'$-invariant.)  {\it Then the form $\Psi^*\sigma$ descends to $\G/\tk$ (i.e. the corresponding function on 
 $\G'$ descends to $\G'/\tk'$). } It is easy to prove this statement in the case when $\tk$ is a Lie algebra of a connected group $\K\subset \G$. Then $\G/\tk$ can be identified with $\G/\K$ and the condition $\psi(k')|\rho\rangle=0$ is equivalent to the condition $|\Psi(K')|\rho\rangle=|\rho|\rangle$ where $K'\in \K'.$ Using $\Psi (g'K')=\Psi(g')\Psi(K')$ we obtain that $ \Psi(g'K')|\rho\rangle= \Psi(g)|\rho\rangle $ hence  $(\Psi^*\sigma)(g'K')= (\Psi^*\sigma)(g')$; this is the fact we need.
 If $\tk$ cannot be represented as a Lie algebra of a group we use similar arguments to prove that the function 
 $ (\Psi^*\sigma)(g')$ is constant on the leaves of foliation of $\G'$  generated by the action of $\tk'$ from the right, hence this function descends to $\G'/\tk'.$
 
 Using the formula (\ref{GGG}) we obtain 
 \begin{equation}\label{PGG}
 (\Psi^*\sigma)(\exp(b)g)=\langle\chi|\Psi(\exp(b)g)|\rho\rangle= \langle\chi|\Psi(\exp(b)\Psi(g))|\rho\rangle.
 \end{equation}
 It is easy to check that $|\Psi(g)\rho\rangle$ descends to $\G/\tk$, therefore
 one can rewrite (\ref{PGG}) in the form
  \begin{equation}\label{PPG}
 (\Psi^*\sigma)(\exp(b)g)= \langle\chi|\Psi(\exp(b)|\rho_P\rangle)
 \end{equation}
 where $g\in\G$ descends to $P\in \G/\tk$ and $\rho_P$  obeys $\tk'_g|\rho_P\rangle=0$ for the subalgebra $\tk_g\subset\tg$ conjugate to $\tk$.
 The formula (\ref{PPG}) can also be applied when $\G$ is a semigroup acting transitively on some space $\P$ with Lie stabilizers  $\tk_g.$
 
 Notice that  if $\delta P=\gamma P$ where $\gamma\in \tg$ then
 \begin{equation}\label{RHO}
 \delta|\rho_P\rangle=\psi(\gamma)\rho_P\rangle.
 \end{equation}
 
 Until now we considered consequences of invariance of $|\rho\rangle$. Similar considerations can be applied to $\langle \chi|$. We obtain
 
{\it  Let us assume that the element $ \langle\chi|\in \mathcal {E'}^*$ is invariant with respect to $\U'\subset \G'$ where $\U$ is a subgroup of $\G.$ Then $\Psi^*\sigma$ is invariant with respect to the left action of $\U'$ on
 $\G'.$ This means  that  $\Psi^*\sigma$ can be regarded as a function on the space of double cosets $\U' \backslash\G'/\tk'$ or as a closed pseudodifferential form on the space of double cosets $\U\backslash\G/\tk.$}


\section {Conformal and superconformal theories}
\subsection {Disks}
The standard conformal disk can be defined as a subset of  $\mathbb{C}$  consisting of points $z\in \mathbb{C}$ obeying $|z|<1.$

The standard NS disk consists of points $(z,\theta)\in \mathbb{C}^{1|1}$ obeying $|z|<1.$
It is equipped with an odd vector field 
\begin{equation}\label {D}D=\partial_{\theta}+\theta\partial_z \end{equation}
The standard R disk consists of the same points. It is equipped with an odd vector field
\begin{equation}\label {DD}D=\partial_{\theta}+\theta z\partial_z \end{equation}
Here $z$ denotes an even coordinate and $\theta$ stands for an odd coordinate.
\subsection {Superconformal manifolds}

One-dimensional complex manifold (=oriented conformal manifold) is pasted together from standard disks by means of holomorphic transformations.

A structure of superconformal manifold on  complex supermanifold of complex dimension
 $(1|1)$  is specified by an odd holomorphic vector field $D$ defined up to multiplication by a non-vanishing holomorphic function. Superconformal transformations are holomorphic transformations preserving $D$ up to a non-vanishing holomorphic factor.  We assume that a superconformal manifold is glued from NS and R standard disks by means of superconformal transformations. This means that in appropriate local coordinates, 
 the field $D$ can be represented either in the form (\ref{D}) or in the form (\ref{DD}).
 \subsection{ Annuli}
 A standard conformal annulus is a part of the standard conformal disk that is singled out  by conditions  
 $0<a<|z|<1.$ In variable $\tau$ related to $z$ by the formula $z=e^{-i\tau}$  it can be represented as a horizontal strip $\log a < {\rm Im}\tau <0$ with identification
 $\tau\sim\tau+2\pi .$
 
 A standard NS annulus and standard R annulus are singled out from standard disks by the same condition. In the coordinates $\tau=i\log z, \zeta=Cz^{\gamma}\theta$ the superconformal structure on annuli is specified by odd vector field 
 $D=\partial_{\zeta}+\zeta\partial_{\tau}$      for an appropriate choice of $C$ and $\gamma$ (for NS annulus we should take $\gamma=\frac 1 2 $, for R annulus   $\gamma=0$          ). The annuli are represented as horizontal strips with coordinates $(\tau, \zeta)$ and identifications $(\tau,\zeta)\sim(\tau+2\pi,-\zeta)$ for NS annulus, $(\tau,\zeta)\sim(\tau+2\pi,\zeta)$  for R annulus.
 
 Imposing the condition $|z|=const$  we obtain NS circle  from NS annulus and R circle from R annulus.
 
 \subsection {Symmetry groups of standard annuli}
 The rotations $z\to e^{-ia}z$ ,where $a$ is real, are symmetries of conformal annulus; they constitute a group $S^1$. In coordinates  $\tau=i\log z$ they have the form  $\tau\to\tau+a$.
 
 The transformations $\tau\to\tau+a$ are also symmetries of standard NS annuli and
 R annuli.  For R annuli we have an additional odd symmetry $\zeta\to \zeta+\alpha.$; the symmetry group is $S^1\times \mathbb{R}^{0|1}.$
 
 In both cases, we have a discrete symmetry $I:(\tau,\zeta)\to (\tau,-\zeta).$ The groups generated by the above symmetries will be denoted by $\Gamma_{NS}$ (for NS annulus)
 and $\Gamma_R$ (for R annulus).
 
 These symmetry groups can be considered also as symmetry groups of NS circle and R circle.
 
 \subsection{Moduli spaces}
 
 Let us denote by $\P_m$ the moduli space of all compact connected one-dimensional complex manifolds ( complex curves) with  $m$ embedded standard conformal disks. In other words, a point of $\P_m$ is defined as a set of $m$ holomorphic embeddings of the standard disk into a compact connected one-dimensional complex manifold. We assume that the images of disks do not overlap, but this assumption is not crucial. The infinite-dimensional space $\P_m$ can be equipped with a complex structure. Pasting together the boundary of the last disk in an element $\P_m$and the boundary of the first disk in an element of $\P_{m'}$ we obtain an element of $\P_{m+m'-2}.$ In other words, we obtain a map $\P_m\times\P_{m'}\to\P_{m+m'-2}$ (sewing map). Similarly,
 pasting together the boundaries of the two last disks we obtain a map $\P_m\to\P_{m-2}.$ Notice, that a more accurate description of these maps (sewing maps) can be given in terms of gluing of standard annuli.
 
 The group $(S^1)^m$ where $S^1$ stands for the symmetry group of the standard conformal annulus acts naturally on $\P_m$; the quotient of $\P_m$ with respect  to  $(S^1)^m$ is denoted by $\hat\P_m.$
 
 Connected components  of $\P_m$ and $\hat\P_m$ are labeled by the genus g of the complex curve; they are denoted  by $\P({\rm g},m)$ and $\hat \P({\rm g},m)$. It is easy to check that the natural map of  $\hat \P({\rm g},m)$ onto the moduli space $\M({\rm g},m)$ of genus g complex curves with $m$ marked points is a homotopy equivalence.

 Instead of manifolds with embedded disks, one can consider complex manifolds with parameterized boundaries.
 
 Sometimes it is convenient to consider two classes of embedded disks: incoming disks and outgoing disks (equivalently incoming boundaries and outgoing boundaries).  If we are using the language of boundaries it is convenient to assume that the orientation of incoming boundaries agrees with the orientation of the manifold and the outgoing boundaries have an opposite orientation. The moduli space of manifolds with $k$ incoming disks and $l$ outgoing disks will be denoted by $\P_k^l$. One can define the sewing  map
 \begin{equation} \label{SEW} \P_k^l\times P_l^r\to \P_k^r\end{equation}
 (outgoing boundaries are pasted to incoming boundaries).
 
 Similarly, we denote by $\P_{m,n}$ the moduli space of all compact connected superconformal manifolds with $m$ embedded standard NS disks and $n$ embedded standard R disks. (The number $n$ is necessarily even.)  This moduli space also can be equipped with complex structure. Gluing
 two standard  NS annuli or two standard R annuli we obtain sewing maps $\P_{m,n}\times \P_{m',n'}\to
 \P_{m+m'-2,n+n'}, \P_{m,n}\to\P_{m-2,n}, \P_{m,n}\times\P_{m',n'}\to \P_{m+m',n+n'-2},\P_{m,n}\to\P_{m,n-2}.$
 The symmetry groups of standard annuli act on $\P_{m,n}$; we define the space $\hat\P_{m,n}$ factorizing $\P_{m,n}$ with respect to this action:
 \begin{equation}\label{PHAT}
 \hat\P_{m,n}=\P_{m,n}/\Gamma_{NS}^m\times\Gamma_R^n.
 \end{equation}
 Again we can consider incoming disks and outgoing disks (or incoming and outgoing NS-boundaries and R-boundaries) . We use the notation $\P_{k,l}^{r,s}$ for the case of $k$ incoming NS disks, $l$ incoming R disks, $r$ outgoing NS disks, $s$ outgoing R disks.

 \subsection{Semigroups of Annuli}
 Let us introduce the notations $$\A=\P_2=\P_1^1, \A_{NS}=\P_{2,0}=\P_{1,0}^{1,0}, \A_R=\P_{0,2}=\P_{0,1}^{0.1}.$$
 It follows from the constructions of the preceding section that these objects can be regarded as semigroups. We say that $\A$ is a semigroup of (conformal) annuli, $\A_{NS}$ is a semigroup of NS annuli, $\A_{NR}$ is a semigroup of R-annuli. All these semigroups can be considered as complex manifolds. The Lie algebra of $\A$ is the Witt algebra
 (the complex Lie algebra diff of complex vector fields on a circle). It is generated by vector fields 
 $L_n=ie^{in\tau}\partial_{\tau}$ obeying
 \begin{equation}\label{WIT}
 [L_m,L_n]=(m-n)L_{m+n}.
 \end {equation}
 The central extension of  Witt algebra is called Virasoro algebra; it is specified by relations
 \begin{equation}\label{VIR}
 [L_m,L_n]=(m-n)L_{m+n}+\frac {c}{12}(m^3-m)\delta_{m+n,0}.
 \end {equation}
where $c$ is called central charge. If $c\neq 0$ a representation of Virasoro algebra can be identified with a projective representation of Witt algebra diff.

 The Lie algebra $W_{NS}$  of $\A_{NS}$ can be interpreted as the complex  Lie algebra of complex vector fields on NS circle (on a super manifold with real even coordinate $\tau$, odd coordinate   $\zeta$ and identification $(\tau ,\zeta)\sim(\tau+2\pi,-\zeta)$). The Lie algebra $W_{NS}$ is generated by even vector fields   $L_n=ie^{in\tau}\partial_{\tau}$ and odd vector fields 
 $G_r=2ie^{ir\tau}\partial_{\zeta}$ where $r$ is half-integer.
 
 The Lie algebra of $\A_{R}$ can be considered as the complex Lie algebra $W_R$ of complex vector fields on R circle $S^1\times \mathbb {R}^{0|1}$ (on a a supermanifold with real even coordinate $\tau$, odd coordinate   $\zeta$ and identification $(\tau,\zeta)\sim(\tau+2\pi,\zeta)$) . It is generated by even vector fields   $L_n=ie^{in\tau}\partial_{\tau}$ and odd vector fields 
 $G_r=2ie^{ir\tau}\partial_{\zeta}$ where $r$ is integer.
 
 In both cases the generators obey
 \begin{equation}\label{NSRWIT}
 [L_m,L_n]=(m-n)L_{m+n},
 [L_m,G_r]=(\frac 1 2 m -r)G_{m+r},
 [G_r, G_s]_+=2L_{r+s}.
 \end {equation}
 One can say that the direct sum of $W_{NS}$ and $W_R$ is a super analog of Witt algebra. Its projective representations correspond to representations of its central extension called super Vrasoro algebra. 

The sewing map $\P_2\times \P_n\to \P_n$ specifies an action of semigroup $\P_2$  and its Lie algebra (Witt algebra)  on $\P_n$
(more precisely, we have $n$ actions of this kind because an annulus can be pasted to any boundary).
Similar statement is true for moduli spaces of superconformal manifolds. (For example, an R-annulus can be pasted to any R-boundary; this generates an action of semigroup  $\A_{R}$ and of corresponding Lie algebra $W_R.$)
 
 \subsection {Conformal theories }
 An axiomatic definition of conformal field theory (CFT) was suggested by Segal and slightly modified in the operator formalism of string theory. In this definition, one assigns a topological vector space  $\H$  to every circle  and a map \begin{equation}\label{SSS} 
 \phi_k^l:\H^k\to\H^l
 \end{equation}
 to every element of $\P_k^l$.(Here $\H^k$ stands for $k$-th tensor power of $\H$.) The sewing map (\ref {SEW}) should agree with the composition of maps (\ref{SSS}.)
 The map $\phi_2^0$ specifies a bilinear inner product in $\H$.

 Alternatively, we can work with moduli spaces $\P_n$ (this is the viewpoint accepted in the operator formalism of string theory).  Then to every point  $P\in \P_n$  we assign an element
 \begin {equation}\label {PHIP}
 |\phi_P\rangle\in \H^n.
 \end{equation}
 where $\H$ is a topological vector space equipped with bilinear inner product.
 This assignment should agree with sewing maps $\P_m\times \P_n\to \P_{m+n-2}$ and $\P_n\to \P_{n-2}$.  To relate this viewpoint with  Segal's definition we identify  $\P_n$ with $\P_0^n$ (all boundaries are outgoing) and use the identification $Hom (\H^k,\H^l)=Hom(\H^{k+l},\mathbb{C})$ where $Hom (A,B)$ stands for the space of continuous linear maps of topological vector spaces $A\to B$.  (This identification follows from $Hom(A\otimes B,C)=Hom(A,Hom(B,C)).$ Notice that our considerations were not rigorous because we never specified which definition of tensor product  we were using and did not specify the topology in the space of maps.) 
 
 The bilinear inner product in $\H$ allows us to work with bra-states $\langle \phi_P|$ (elements of the dual space) instead of ket-states (\ref{PHIP}). (An element of $\H$ specifies a linear functional on $\H$ as inner product with this element.)
 
 Notice that in the original Segal's definition, it is assumed that the maps (\ref{SSS}) belong to the trace class. This assumption is appropriate for theories describing fields taking values in compact manifolds. As was noticed in the published version of \cite{SEG} in general case one should not assume that the operators (\ref {SSS}) have trace; these operators should be defined only if $l\geq 1$ (there exists at least one outgoing boundary). In particular, the case $k=l=0$ should be excluded hence the partition function of CFT cannot be defined rigorously. Nevertheless one can define the partition function using appropriate regularization and to prove that  it vanishes in superconformal field theory
 that appears in string theory with space-time supersymmetry. The inner product in $\H$ can be defined only on dense subset, in this case only for a part of elements of $\H$ (ket-states) we can define a bra-state. (Notice that we do not assume that $\H$ is a Hilbert space. In concrete situations one can use different definitions of $\H$; we do not pay attention to these subtleties.)
 
 In the definition above it is assumed that the central charge $c=0.$ For non-zero central charge the maps (\ref{SSS}) are projective ( the image of every element is defined up to a constant factor).
 
 As we know, the semigroup of annuli $\A$ acts on moduli spaces and this action induces the action of Witt algebra diff.  These actions induce representations  of $\A$ and diff
  in $\H$. Integrating the representation of diff in $\H$ we obtain a representation of $\A$ in $\H$.
  
   If maps (\ref{SSS}) are projective 
 these representation are also projective.
 
 To  construct CFT  we can start with classical two-dimensional theory specified by local conformally invariant action functional $S$, 
 (The functional $S$ can depend on Riemannian metric $g_{ab}(x)$ but it should not change when we multiply the metric by a non-vanishing function.) For example, for scalar field $\Phi$ we  take
 \begin{equation}
 \label {SCAL}
 S=\int g^{ab}\partial_a\Phi\partial_b\Phi\sqrt gd^2x
 \end {equation}
 Let us consider a functional $S$  on fields defined on the standard annulus. The variation $\delta S$ of the functional $S$ can be regarded as one-form on the space of fields; it can be represented as as a sum of bulk term vanishing on fields satisfying equations of motion and two boundary terms $\alpha_+$ and$-\alpha_-$. We define a phase space $H$ as the space of solutions of equations of motion
 on annulus equipped with with symplectic form $\omega$ defined by the formula $\omega=\delta \alpha_+=\delta \alpha_-$. ( More precisely, we should consider solutions of equations of motion on infinitesimally thin annulus. We denote by $\delta$ the de Rham differential on the space of fields.   The relation  $\delta \alpha_+=\delta \alpha_-$ on the phase space follows from 
 $\delta^2S=0$. The two-form $\omega$ is closed; we assume that it is non-degenerate, then it specifies a symplectic structure.)
 
  Quantizing the phase space we obtain the space $\H$ of CFT. Now let us take a point  $P\in \P_n$ considered as two-dimensional conformal manifold with parametrized boundaries. The the space of solutions of equations of motion on $P$ specifies a submanifold $L_n$ of a product of $n$ copies of phase space $H$ (we restrict every solution to  neighborhoods of  boundary circles). It follows from Stokes' theorem  that  $L_n$ is isotropic (the symplectic two-form vanishes on $L_n$).{\it  We assume that it is Lagrangian}, (i.e. it cannot be extended to larger isotropic manifold). Then we can quantize it to get  a point ( more precisely, a ray) in the space $\H^n$. This procedure is ambiguous  but we can hope that with a right choice of the quantization procedure we obtain CFT (in general with non-zero central charge).
  
  It is important to notice that in the case of quadratic action functional (=linear equations of motion) this procedure can lead to rigorous constriction of CFT (at least in case when the quadratic action functional is non-degenerate).
    
  For quadratic action functional  the space of solutions of equations of motion is a linear space, in particular, the phase space $H$  is a linear space. Poisson brackets of fields are constant. After quantization we  obtain an algebra with generators  $\hat w_n$ corresponding to elements $w_n$ of the basis of $H$. The commutators of generators $\hat w_n$ are constants; for a right choice of basis the operators $\hat w_n$ obey canonical commutation relations. Let us denote by $a_n$ the basis of Lagrangian submanifold $L_1\subset H$ and by $\hat a_n$ the corresponding generators.(Here we consider $L_1$ as a space  of solutions of equations of motion on a sphere with one hole.). The corresponding vector $\Phi\in\H$ obeys $\hat a_n\Phi=0$. (This condition specifies $\Phi$ up to a numerical factor.) Extending the basis
  $a_n$ of $L_1$ to a basis in the phase space $H$ we can say that $\H$ can be regarded as Fock representation of Weyl algebra 
  $\mathcal {W}$ with generators $\hat a_n^^+, \hat a_n$ obeying CCR (canonical commutation relations). Notice, however, that in this construction $\hat a_n$ and $\hat a^+_n$ are not necessarily Hermitian conjugate and the vector $\Phi$ is not necessary normalizable, hence in general $\H$ is not the standard Fock space. (Recall that we do not assume that $\H$ is a Hilbert space; saying that a vector is not normalizable we have in mind that $\H$ cannot be equipped with a structure of Hilbert space in such a way that the vector is normalizable. A vector corresponding to a real Lagrangian submanifold is always non-normalizable.)
  
  To quantize  $L_n$ considered as linear Lagrangian submanifold of $H^n$ we notice that the corresponding vector $\Phi_n\in \H^n$ obeys $\hat A\Phi_n=0$ for every $A\in L_n$. ( Here $\hat A$ stands for the element of the direct product of $n$ copies of Weyl algebra $\mathcal {W}$ that corresponds to $A\in L_n$.) This condition specifies $\Phi_n$ up to a numerical factor.  
  
  The vector (or, more precisely the ray) $\Phi_n$ should be identified with the vector $|\phi_P\rangle$ of operator formalism. One can use CTA ( charge transfer argument ) of \cite {A} to check that it satisfies sewing condition up to a factor, hence we obtained CFT
  (in general with non-vanishing central charge). (In the case when the inner product is defined only on dense subset of $\H$ and we cannot identify $\H$ with its dual we need additional assumptions to make CTA rigorous.)

   Let us consider as an example a real scalar field with action functional (\ref {SCAL}). The solutions of equations of motions
  are harmonic functions. On the standard annulus every harmonic function can be represented  as a sum of holomorphic function
  $f=\sum _{n\neq 0}c_nz^n$ , complex conjugate antiholomorphic function $\bar f$ and linear combination of functions $1$  and $\log |z|$.
  
  The above procedure leads to CFT with central charge $1.$ We will always neglect the contribution from $(1, \log |z|)$. Than  we can say that this CFT splits into tensor product of
   holomorphic(=chiral=left-handed) theory and antiholomorphic (=antichiral=right-handed) theory. This is a general feature of two-dimensional CFT.
  If antiholomorphic part is complex conjugate to holomorphic part (as for scalar field) we say that CFT is symmetric.
  
  In the space $\H$ of general CFT we have two actions of Virasoro algebra
  (one comes from chiral sector, another from antichiral sector). In symmetric CFT these actions are complex conjugate.
  
  Let us define a field of type $(p,q)$ as a field transforming as $(dz)^p(d\bar z)^q$ by a holomorphic change variables. A holomorphic field transforming as $(dz)^p$
  (a field of type $(p,,0)$) is called $p$-differential. A product of $J$- differential and $
(1-J)$-differential can be integrated over a real curve. If $D_J$ stands for the space of $J$ differentials on conformal manifold $X$ then this construction specifies a bilinear form $\int_{\Sigma} bc$. Here $\Sigma$ stands for a closed 
  curve in $X$  (only homology class of $\Sigma$ is relevant) and $b\in D_J$,$c\in D_{1-J}.$  If $X$ is an annulus then antisymmetrization of this bilinear form  gives  symplectic structure on the direct sum $D_J+D_{1-J}.$ Quantizing this symplectic manifold (=using CCR for $b,c$) and  considering elements of  $D_J+D_{1-J}$ as solutions
  of equations of motion for every conformal manifold we obtain CFT called
  $(\beta,\gamma)-$ model.  Symmetrizing the same bilinear form we obtain a structure of odd symplectic manifold on  $D_J+D_{1-J}$ for annulus. Quantizing
  it (=using CAR for $b,c$) we obtain CFT called $(b,c)$-model. (Notice that in
  $(b,c)$-model we can work with normalizable vectors but in $(\beta,\gamma)$-model we should consider non-normalizable vectors.) In similar way working with antiholomorphic fields we obtain $(\tilde \beta,\tilde \gamma)$-model
  and $(\tilde b,\tilde c)$-model.
  
  We assume that $J$ is an integer in these constructions. They can be applied also for half-integer $J$, however, {\it we do not get genuine CFT in this case} (to define
  $J$-differential with half-integer $J$ we need a spin-structure on conformal manifold). This remark is important in superstring theory.
  
  Notice, that one can obtain any element of $\P_n$ pasting together annuli (=spheres with two  holes) and standard spheres with  three holes (= round spheres with three round holes). Using this remark  we can check that one can 
  restore CFT knowing two-point and three-point correlation functions as well as the action of Virasoro algebra on $\H.$ ( One can define correlation functions on a sphere taking a sphere with $n$ round holes with radii tending to zero.)
   
 \section{Superconformal theories}
   In the axiomatic description of superconformal theory (SCFT) with zero central charge, one assigns a topological 
 $\mathbb {Z}_2$-graded vector space $\H_{NS}$ to every NS-circle, a topological   $\mathbb {Z}_2$-graded  vector space $\H_R$ to every R-circle and a map 
 \begin{equation}\label {SNR}
 \H_{NS}^k\otimes \H_R^l\to\H_{NS}^r\otimes \H_{R}^s
 \end{equation}
 to every element of 
  $\P_{k,l}^{r,s}$ (to every superconformal manifold with $k$ incoming NS disks, $l$ incoming R disks, $r$ outgoing NS disks, $s$ outgoing R disks). Again the sewing map should agree with the composition of maps (\ref{SNR}).
  
  Alternatively, to every superconformal manifold  $P$ with $m $ NS disks and $n$  R disks  (to every element of $\P_{mn}$) we can assign an element 
  \begin{equation}\label {PHIS}
  |\phi_P\rangle\in \H_{NS}^m\otimes \H_R^n
  \end{equation}
  and require compatibility with sewing maps and inner products in $\H_{NS}$ and $\H_R$.
  
  An example of superconformal theory can be obtained from action functional
  \begin{equation}\label {SUCO}
  S=\int D X\bar DXdzd\bar z
  \end {equation}
  where $X$ is a real field on a superconformal manifold. In  local coordinates $(z,\theta)$ we have
  $D=\partial_{\theta} +\theta\partial_z, X=\Phi (z,\bar z)+\theta\psi(z,\bar z)+\bar\theta\bar\psi(z,\bar z)+\bar \theta\theta K(z,\bar z).$
  Solutions of corresponding equations of motion are superharmonic functions (functions obeying $\bar D DX=0)$.
  It is easy to check that for superharmonic function $X$ the function $\Phi$ is harmonic, the odd function $\psi$ is holomorphic and $K=0$. This means that our theory considered as CFT describes scalar boson $X$, left-handed fermion $\psi$ and right-handed fermion $\bar\psi$. ( As usual we can consider chiral and anti-chiral parts of the theory.) 
  
  We quantize the theory using CCR for boson and CAR for fermions.  The quantized theory can be regarded as tensor product of chiral (left-handed) and antichiral (right-handed) parts. Two copies of super Virasoro algebra act in $\H$ (The generators of left one are denoted by $L_n,G_n$, the generators of right one are denoted by $\tilde L_n,\tilde G_n$.)
  
  Notice that  the fermion $\psi$ can be regarded as $\frac 1 2$-differential.
  As we emphasized in the definition of  $\frac 1 2$-differential we need spin-structure therefore strictly speaking we do not obtain CFT quantizing $\psi$, hence we do not get SCFT quantizing   (\ref {SUCO}). However, we can obtain SCFT from (\ref{SUCO}) noticing
  that our theory is invariant with respect to the involution $I$  that changes the sign of
  odd coordinate $\theta$.   The involution $I$ can be lifted to $\H$; more precisely we obtain two involutions $I_{left}$ and $I_{right}$  (the involution $I_{left}$ transforms $G_n$ into $-G_n$, the involution $I_{right}$ transforms $\tilde G_n$ into $-\tilde G_n$). The involutions $I_{left}$ and $I_{right}$ can be chosen in different ways; this choice specifies the type of superstring. (In particular, the type $II_A$ superstring is left-right symmetric.)
  
  \subsection{ Topological (super)conformal theories}
  
   In the definition of topological conformal field theory (TCFT),  the space $\H$ is equipped with a differential. The map (\ref{SSS}) is assigned to every element of ${\P'}_k^l$; it should agree with differentials. The sewing map 
  \begin{equation} \label{TSEW}{ \P'}_k^l\times{\P'}_l^r\to {\P'}_k^r\end{equation}
  should agree with the composition of maps (\ref{SSS}).
  
  In operator formalism we should consider maps $P'_n\to \H^n$ agreeing with sewing maps and differentials.
  
  In the definition of topological superconformal theory (TSCFT) we should assume that the spaces $\H_{NS}$ and $\H_R$ are equipped with differentials and the maps (\ref{TSEW}) assigned to elements of    ${\P'}_{k,l}^{r,s}$ agree with differentials. The maps (\ref{TSEW}) should agree with sewing maps.
  
  Notice that the above definitions of TCFT and TSCFT are slightly different from definitions given in previous papers (we consider 
   functions on a space $\P'$, i.e. pseudodifferential forms on $\P$, previous definitions are formulated in terms of forms on $\P$; see \cite{BEL} for details).{\footnote {The paper \cite {BEL} is based on the notion of differential $r|s$- form on supermanifold introduced by Voronov and Zorich \cite{VZ} and related to pseudodifferential forms by Baranov-Schwarz transform \cite {BS}.
   We do not use the notion of $r|s$-form; this leads to essential simplifications.}}

   The most important examples of TCFT are constructed from  CFT with central charge $c=26$ (matter CFT): the space $\H$ is defined as the tensor product of the space of states of matter CFT  by the space of ghosts (by spaces of $bc$-model and $\tilde b\tilde c$-model with $J=2$).  The most important examples of TSCFT can be constructed from SCFT with critical central charge in a similar way.
  
 \section{String amplitudes}
\subsection {String amplitudes for closed bosonic string}

 For critical bosonic string, we obtain one-string space of states $\mathcal{E}'$ adding ghosts to the matter sector ( to CFT with central charge $=26$).We have the action of Virasoro operators 
  $\hat L_n,  \tilde {\hat L}_m$ , ghost operators $ b_n, \tilde b_n$ and BRST operators $Q,\tilde Q.$  (The Virasoro operators are obtained as sums of Virasoro  operators of matter sector and ghost sector. They have central charge equal to $0$.) We use the following relations between these  operators 
   \begin {equation}\label {TCFT}
   \begin {split}
   [\hat L_m,\hat L_n]=(m-n)\hat L_{m+n}\\
    [\tilde {\hat L}_m,\tilde{\hat  L}_n]=(m-n)\tilde{\hat L}_{m+n}\\
     [\hat L_m, b_n]=(m-n)b_{m+n}, [b_m,b_n]_+=0\\
      [\tilde {\hat L}_m, \tilde b_n]=(m-n)\tilde b_{m+n}, [\tilde b_m,\tilde b_n]_+=0\\   
   \hat L_n=[Q, b_n],  \tilde {\hat L}_n=[\tilde Q, \tilde b_n], [Q,Q]_+=0, [\tilde Q,\tilde Q]_+=0
   \end{split}
   \end {equation}
In other words, we have two representations of the complex Lie algebra diff$''$. These representations commute, i.e. we have a representation of 
 the Lie algebra diff$''+$diff$''$. 
 
 The group $(S^1)'$ where $S^1$ stands for the symmetry group of the standard annulus acts on $\mathcal{E}'$; generators of its Lie algebra are $L_0-\tilde L_0, b_0-\tilde b_0.$ We say that vector $|\chi\rangle\in \mathcal{E}'$  belongs to $\hat{\mathcal{E}}$ if it is ($S^1)'$-invariant (i.e. obeys   $(L_0-\tilde L_0)|\chi\rangle=0, (b_0-\tilde b_0)|
 \chi\rangle=0$). In what follows instead of elements $|\chi\rangle\in \ \hat{\mathcal{E}}$ we will consider corresponding elements of the dual space
 denoted by $\langle\chi|$.

 The cohomology of BRST-operator $Q+\tilde Q$ in the space $\hat{\mathcal{E}}$ is called semi-relative cohomology. Classes of semi-relative cohomology correspond to particles (to scattering states).
 
We will apply the considerations of Section 2 to the diagonal part $D''$ of the Lie algebra  diff$''+$diff$''$. We fix a compact complex curve $P_0$ of genus g with $n$ embedded disks and define a subalgebra $\tk$ of the Lie algebra $D^n$ as a vector field on the union of boundaries of the disks (=a set of vector fields on boundaries of the disks) that can be extended holomorphically to the complement of the disks. The complex Lie algebra $D$ can be regarded as a Lie algebra of the semigroup $\A$ of conformal annuli considered as complex manifold, the curve $P_0$ specifies an element of the moduli space $\P_n$ and $\tk$ can be interpreted as the Lie stabilizer at the point $P_0\in \P_n$ of the action of the Lie algebra $D^n$ of the semigroup $\A^n$ on $\P^n.$ The semigroup $\A^n$ acts transitively on the connected component $\P({\rm g},n)$ of $\P_n$; hence we can identify $\A^n/\tk$ with $\P({\rm g},n)$ (the space $\P({\rm g},n)$  consists of  elements of $\P_n$ represented by genus g conformal manifolds).
 
 We will apply the considerations of Section 2 to obtain a class of physical quantities that contains string amplitudes.
 Namely, we will consider an element $\sigma$ specified by the formula (\ref{SIG}) where $|\rho\rangle=|\phi_{P_0}\rangle .$  ( Recall that the state corresponding to $P\in \P_n$ in operator formalism is denoted by  $|\phi_P\rangle\in (\mathcal{E}')^n$.)
  It follows from the axioms of CFT that
 $|\phi_{P_0}\rangle$ is $\tk'$-invariant. This means that   the form $\Psi^*(\sigma)$ descends to $\A^n/\tk=\P({\rm g},n)$. 
 
 We will use (\ref {PPG}) taking $\G=\A^n$ and $|\rho_p\rangle=|\phi_p\rangle$ ( to identify $\rho_P$ and $\phi_p$ we can use (\ref{RHO})).The  formula (\ref{PPG}) allows us
 to show that $(\Psi^*\sigma)(\exp(b)g)$ where $g\in \A^n$ and $b\in \Pi D^n$ descends  to $P\in \P({\rm g},n)$ 
  in the following way:
  \begin{equation}\label{PPGG}
 (\Psi^*\sigma)(\exp(b)g)= \langle\chi|\Psi(\exp(b))|\phi_P\rangle
 \end{equation}
 The formula (\ref {PPGG}) specifies a closed inhomogeneous differential form on $ P({\rm g},n)$. Imposing some conditions on $\chi$ we get a form descending further to the space $\hat \P({\rm g},n)$ that can be obtained from $ \P({\rm g},n)$ by means of factorization with respect to $(S^1)^n$.  The homology of $\hat \P({\rm g},n)$  can be identified with the homology of the moduli space  $\hat \M({\rm g},n)$ of genus g curves with $n$ marked points, hence we can integrate the forms on  $\hat \P({\rm g},n)$ over cycles in   $\hat \M({\rm g},n).$
 Integration over the fundamental cycle in   $\hat \M({\rm g},n)$ leads to string amplitudes. (This statement suffers  from the standard problems of  string theory: the space   $\hat \M({\rm g},n)$ is non-compact, the fundamental cycle is  locally finite, but not finite, and the integral is diverging.) The conditions on $\chi$ that we mentioned are the standard conditions imposed in the closed bosonic string theory on scattering states, namely $ \chi$ should be $((S^1)^n)'$-invariant. 
 If the state $\chi$ is represented as a tensor product of states $\chi_k$ corresponding to different components of the boundary this means that
  the operators $L_0-\tilde L_0$ and $b_0-\tilde b_0$ corresponding to the $k$-th component of the boundary should annihilate $\chi_k$ (hence $\chi_k$ should belong to $\hat{\mathcal{E}}$).
  
  To prove that the form $\Psi^*\sigma$ descends to $\hat \P({\rm g},n)$ for $((S^1)^n)'$-invariant $\chi$ we use the considerations at the very end of Section 2 and the remark that  $\hat \P({\rm g},n)$ can be identified with the space of double cosets $(S^1)^n\backslash \A^n/\tk.$
  
  To verify that our formulas lead to string amplitudes we recall that our $\phi_P$ coincides with a state denoted in the same way in operator formalism and apply the formula (\ref{INT}) for the integral of pseudodifferential form. ( We use the fact that in operator formalism string amplitudes are obtained by means of integration of the expression $\langle\chi  |B|\phi_P\rangle $ where $B$ is a polynomial with respect to $b_k,\tilde b_k$.)
  
  \subsection{Scattering amplitudes for superstring}
  
  The situation in the case of superstring is very similar. We obtain the one-string space $\mathcal{E}'$ by adding ghosts to the matter sector. The Virasoro operators $L_n, \tilde L_n$ and supersymmetry operators $G_r,\tilde G_r$ together with ghost operators $b_n,\tilde b_n$, $\beta_n,\tilde \beta_n$ and BRST operators define representations of two copies of the Lie algebra $W''$ in one-string space. These representations specify a representation of the direct sum of two copies of 
 $W''$. (Recall that $W$ is by definition a direct sum of Neveu-Schwarz algebra $W_{NS}$ and Ramond algebra $W_R$.)  
 
 One can define an action of the group $\Gamma'_{NS}\times\Gamma'_R$ on one-string space. As in the bosonic case, the generators of the Lie group $(S^1)'$ act as $L_0-\tilde L_0$ and  $b_0-\tilde b_0$. The elements of Lie algebra corresponding to the odd symmetry $\zeta\to \zeta+\alpha$ are defined as 
 $G_0-\tilde G_0$ and $\beta_0-\tilde \beta_0.$ To define a transformation  corresponding to the discrete symmetry   we use relations $IG_r=-G_rI, I\tilde G_r=-\tilde G_rI.$ (In non-symmetric case we lift the discrete symmetry to two different operators in $\H$ denoted by $I_{left}$ and $I_{right}$. They obey $I_{left}G_r=-G_rI_{left}, I_{right}\tilde G_r=-\tilde G_rI_{right}.$) The invariance with respect to  $I_{left}$ and $I_{right}$ is equivalent to GSO projection \cite{BEL}.
 
 We say that $|\chi\rangle\in  {\mathcal{E}'}$ belongs to  $\hat{\mathcal{E}}$ if it is   $\Gamma'_{NS}\times\Gamma'_R$-invariant.

 Notice that we have some freedom in the construction of one-string space (one can use different pictures in $(\beta,\gamma)$-model) and in the definition of discrete symmetry.

 We fix a compact connected superconformal manifold $P_0$ with $m$ embedded NS disks and $m$ embedded R disks. It can be regarded as a  point of the supermoduli space $\P_{m,n}$.  The direct product $\A_{m,n}$  of $m$ copies of semigroup $\A_{NS}$ and $n$ copies of semigroup $\A_R$ acts on $\P_{m,n}$; the action of this semigroup induces an action of its Lie algebra  $W_{NS}^m\otimes W_R^n$. We define a Lie subalgebra $\tk$ of this  Lie algebra as a Lie stabilizer at the point $P_0\in P_{m,n}$.
 
 Now we apply the constructions of Section 2  taking  $\A_{m,n}$ as $\G$. We take as $\sigma$ an element specified by the formula (\ref{SIG})
 where $|\rho\rangle$ belonging to the space  $(\mathcal {E}')^{m+n}$  is $\tk'$-invariant. We assume that $|\chi\rangle\in (\mathcal {E}')^{m+n}$ ) can be represented as a tensor product of vectors $|\chi_k\rangle\in \hat{\mathcal{E}}.$
 
 To prove that this construction leads to string amplitudes we notice
 that the semigroup $\A_{m,n}$
 acts transitively on the connected component $\P({\rm g}, m,n)$ of $\P(m,n)$(on a part of the moduli space $\P(m,n)$ consisting of superconformal manifolds having the same genus g as $P_0.$). This allows us to identify
   $\A_{m,n}/\tk$ with $\P({\rm g}, m,n)$.  The group $SYM (m,n)=(\Gamma_{NS})^m\times (\Gamma_R)^n$ acts on  $\P({\rm g}, m,n)$; we denote  the space of orbits of this action by $\hat\P({\rm g}, m,n).$ One can identify the space  $\hat\P({\rm g}, m,n)$ with the space of double cosets $SYM(m,n)\backslash \A_{m,n}/\tk.$ This remark allows us to say that the pseudodifferential form $\Psi^*(\sigma)$ descends to   $\hat\P({\rm g}, m,n)$ if $|\rho\rangle$ in (\ref{PS}) is $\tk'$-invariant and $|\chi\rangle$ is $SYM(m,n)'$-invariant (we use the considerations at the end of Section 2). Notice that we can take $|\chi\rangle$ as a tensor product of elements $\chi_k\in \hat{\mathcal{E}}$.  

In the operator formalism, we have states $|\phi _P\rangle\in (\mathcal{E}'_{NS})^m\otimes(\mathcal{E})'_R)^n$ coming from central charge zero SCFT
combining matter and ghosts (this SCFT can be regarded as TSCFT). Taking $|\rho\rangle=|\phi_{P_0}\rangle$  we obtain that $|\rho|\rangle$ is $\tk'$-invariant hence we can apply the constructions above. We obtain string amplitudes by integrating $\Psi^*(\sigma)$. This follows from comparison with formulas in \cite{BE},\cite{BEL},\cite{BELO}. (The space $\hat\P$ is defined in these papers in a different way: our definition of $\hat\P$ contains additional factorization with respect to discrete group. This factorization is related to summation over spin- structures.)

We did not use the ghost number in our arguments. However, as we have emphasized we can extend $\tg''$ including an even generator corresponding to the ghost number. Considering  the ghost number we can check that many amplitudes we defined vanish. It was stressed in \cite {WIT} that we need the ghost number to rigorously define integrals appearing in the $\beta\gamma$- model.

Notice, that in our considerations as well as in the analysis of bosonic string we tacitly assumed that the theory is left-right symmetric (as for type $II_A$ superstring).  For heterotic string, we have  TCFT in the left-moving sector and TSCFT in the right-moving sector.
The above considerations can be applied separately to both sectors. The integration cycle can be chosen as in the conventional approach (see\cite{WIT}). The type $II_B$ superstring also is not left-right symmetric; again we apply our considerations separately to both sectors.

{\bf Acknowledgments} I am indebted to L. Alvarez-Gaume, M.Movshev, A.Rosly, A. Tseytlin, E.Witten, and A. Zeitlin for useful discussions and to anonymous referee for his remarks.
I appreciate the hospitality of IHES,IAS, Simons Center, Caltech and M.Kontsevich, E. Witten, L. Alvarez-Gaume, N. Nekrasov, A.Kapustin.

This work was supported by NSF grant PHY-2207584.

\end{document}